\documentstyle[twocolumn,aps,epsfig]{revtex}

\begin{document}
\title{Is the existence of a softest point in the directed flow excitation function\\
an unambiguous signal for the creation of a quark-gluon plasma?}

\author{Marcus Bleicher and J\"org Aichelin}

\address{ SUBATECH,
Laboratoire de Physique Subatomique et des Technologies Associ\'ees \\
University of Nantes - IN2P3/CNRS - Ecole des Mines de Nantes \\
4 rue Alfred Kastler, F-44072 Nantes Cedex 03, France}

\maketitle

\noindent
\begin{abstract}
The excitation function of the in-plane directed flow of nucleons is studied
within a non-equilibrium transport approach.
It is demonstrated that a local minimum in the excitation 
function of the directed flow develops, which is not related to a transition
into a quark-gluon plasma (QGP) phase. 
It is a consequence of the dynamical softening of the underlying equation 
of state, due to the onset of resonance matter and particle production.
Thus, the interpretation of this minimum as a 'smoking gun' signature for the
creation of a QGP is premature. 
\vspace{.6cm}
\end{abstract}

The excitation functions of hadron ratios and hadron flow has since long been
suggested to search for evidence of exotic states and phase transitions
in nuclear collisions~\cite{Rischke:1995pe,qgpreviews}.

Especially the in-plane collective flow is
the earliest predicted observable to probe heated and 
compressed nuclear matter~\cite{scheid68a}.
Its sensitivity to the equation of state (EoS) might be used to search 
for abnormal matter states and phase 
transitions~\cite{hofmann76a,stocker:86,Stocker:1979mj}.
Until now, the study of in-plane and azimuthal flow in high 
energy nuclear collisions is attracting large attention from both
experimentalists and theorists~\cite{Csernai:1982ff,H-G,E877,liu97,posk98}. 

In fluid dynamics, the transverse collective flow is intimately 
connected to the pressure $P(\rho, S)$  (which in turn depends on 
the density $\rho$ and the entropy $S$) of the matter  in the reaction
zone \cite{stoecker81a}:
\begin{equation}
\label{pxeqn}   
{\bf p} \,\propto\, \int_t \int_A P(\rho,S) \, {\rm d}{\bf A} \, {\rm d}t\,.
\end{equation}
Here d{\bf A} represents the surface element between the participant and
spectator matters and the total pressure is the sum of the
potential pressure and the kinetic pressure, while $t$ denotes the time over which
the pressure acts. Thus, the transverse 
collective flow depends directly on the equation of state, $P(\rho,S)$.

The sensitivity of the flow on the equation of
state~\cite{Csernai:1982ff,H-G,sorge99,soff,bravina,Kolb:2000sd} which governs 
the evolution of the system  created in violent nucleus-nucleus collisions is conveniently
addressed in terms of Fourier coefficients $v_i$ of the azimuthal distribution 
(given by the angle $\phi$) of the explored hadrons. At fixed rapidity on expands:
\begin{equation}
\frac{{\rm d}N}{{\rm d}\phi}= 1 + 2\, v_1\, {\rm cos}(\phi) + 2\, v_2 \,{\rm cos}(2\phi) \quad,
\end{equation}
with 
\begin{equation}
v_1 = \left\langle\frac{p_x}{\sqrt{p_x^2+p_y^2}}\right\rangle\quad,\quad 
v_2 = \left\langle\frac{p_x^2-p_y^2}{p_x^2+p_y^2}\right\rangle\quad.
\end{equation}

The first coefficient describes the directed in-plane flow. The directed flow is most 
pronounced in semi-central interactions around target and projectile rapidities 
where the spectators are deflected away 
from the beam axis due to a bounce-off from the compressed and heated matter in the overlap region.
The time scales probed by the directed flow are set by the crossing time of the Lorentz-contracted
nuclei. Thus, it serves as keyhole to the initial, probably non-equilibrium, stage of the reaction.
 
In contrast, the elliptic flow~\cite{sorge99,olli92,lshur,risc96,sorge97,heisel,csernai,brachmann,zhang99,Bleicher:2000sx}  
as measured by the $v_2$ coefficient is generated after the overlap of the initial nuclei. 
This type of flow is strongest around central rapidities in semi-peripheral collisions. It is 
driven by the anisotropy of the pressure gradients, due to the geometry of the initial overlap region.
Therefore, it is a valuable tool to gain insight into the expansion stage of the fire ball

Both types of flow can be used to investigate the so-called 'softest point' - i.e. a 
local minimum of $P/\epsilon$ as a function of the energy density $\epsilon$ - in the EoS \cite{hung}.
As pointed out in \cite{hung,test1}, the existence of this 'softest
point' leads to a prolonged expansion of matter and consequently to a
long lifetime of a mixed phase of QGP and hadron matter.
Analogously, it also takes longer to compress matter in the early stage
of a heavy--ion collision \cite{test2}. These features result in two key predictions as
'smoking gun' signatures for Quark-Gluon-Plasma formation:
\begin{itemize}
\item A {\it kinky} centrality dependence of the scaled elliptic flow~\cite{Sorge:1998mk} and
\item  a {\it minimum} in the excitation function of the directed 
in-plane collective flow \cite{Rischke:1995pe}.
\end{itemize}
Especially, the observation of this local minimum in the energy dependence of 
the in-plane directed flow
\begin{equation}
p_x^{\rm dir} = \frac{1}{M} \sum_i^M \, p_{x,i} \, {\rm sgn}(y_i)
\end{equation}
in heavy--ion collisions - here $i$ sums over all considered nucleons, 
$p_{x,i}$ is the momentum in $x$ direction of nucleon $i$ and $y_i$ denotes the
rapidity of nucleon $i$ - has been suggested to be a clear signature 
for a change of nuclear matter properties
as for instance in the transition from hadron to quark and gluon degrees
of freedom (see also the pioneering works by~\cite{hofmann76a,csernai2,bravina2}). 

Recently, a more complex dynamical picture behind the softening of the directed flow
had been discovered. In Refs. \cite{brachmann} and \cite{Csernai:1999nf} it
was argued that in case of a QGP creation at first strong directed flow develops, which is
later compensated by an anti-flow of nucleons at central rapidities.

In this letter, we study the excitation function of the 
directed flow and its connection to the underlying equation of state. 
We demonstrate that the common interpretation of the softening of the equation of 
state as a unique signature of a mixed phase near a transition to a QGP
is premature.
 
For our investigation, the Ultra-relativistic Quantum Molecular 
Dynamics model (UrQMD 1.2)~\cite{urqmd} is applied 
to heavy ion reactions from $E_{\rm beam}= 100$~AMeV to 500~AGeV. 
This microscopic transport approach is based on the covariant propagation of
constituent quarks and diquarks accompanied by mesonic and baryonic 
degrees of freedom. It simulates multiple interactions of 
ingoing and newly produced particles, the excitation
and fragmentation of color strings and the formation and decay of
hadronic resonances. 
Towards higher energies, the treatment of sub-hadronic degrees of freedom is
of major importance.
In the present model, these degrees of freedom enter via
the introduction of a formation time for hadrons produced in the 
fragmentation of strings \cite{andersson87a,andersson87b,sjoestrand94a}.
The leading hadrons of the fragmenting strings contain the valence-quarks 
of the original excited hadron. In UrQMD they are allowed to
interact even during their formation time, with a reduced cross section
defined by the additive quark model, 
thus accounting for the original valence quarks contained in that
hadron \cite{urqmd}. A phase transition to a quark-gluon state is 
not incorporated explicitly into the model dynamics. However, 
a detailed analysis of the model in equilibrium, yields an effective equation of state of 
Hagedorn type \cite{Belkacem:1998gy,Bravina:1999dh}.

Note that the present simulations have been performed without 
potential interactions. Thus the absolute magnitude of the directed flow at the 
lowest energies is certainly underpredicted~\cite{Cugnon:nk}. 
Nevertheless, the general statement of a non-monotonic dependence of the 
directed flow on the collision energy is not blured by this detail. 

As a measure for particle production, we define an inelasticity:  
\begin{equation}
{\rm Inelasticity}=\frac{\sum\, m_i}{E_{\rm total}}\quad 
{\rm at\, y_{\rm cm}}\pm 0.5\quad,
\end{equation}
the sum is taken over all newly produced particles, 
with $m_i$ being the mass of each produced particle and $E_{\rm total}$
being the total available energy, $E_{\rm total}=A\sqrt s$, for a given 
center-of-mass energy $\sqrt s$.

\begin{figure}
\vspace*{-.5cm}
\par \centerline{\resizebox*{!}{0.5\textheight}
{\includegraphics{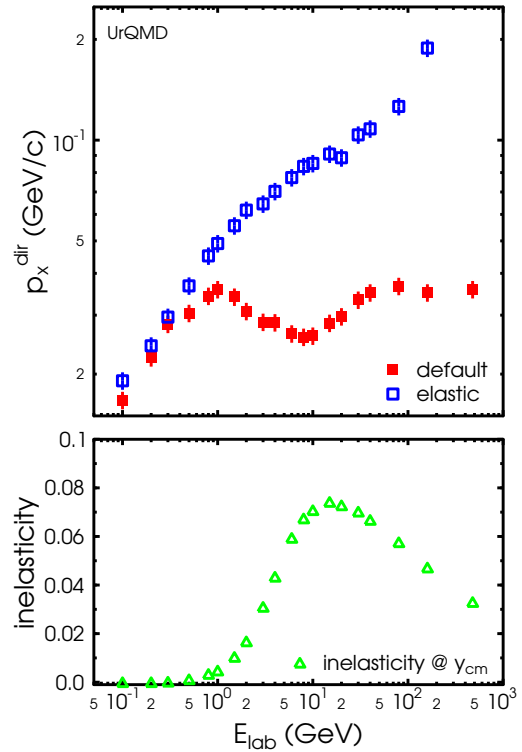}}} \par{}
\caption{Excitation functions for central Au+Au (Pb+Pb) reactions.
Top: Directed flow $p_x^{\rm dir}$ 
of nucleons with only isotropic elastic interactions (open squares)
and with full elastic and inelastic collision term (full squares).
Bottom: Inelasticity (open triangles).
\label{exc.pxdir}}
\end{figure}

Fig. \ref{exc.pxdir} depicts the excitation functions of the directed flow 
of protons (full squares) and the inelasticity (open triangles) 
for central Au+Au (Pb+Pb) reactions with b$\le 3.4$~fm.
One clearly observes a non-monotonic behavior of the directed flow
excitation function from UrQMD: Up to 1~AGeV beam energy the directed flow 
of nucleons increases. As particle production sets in, the directed 
flow starts to decrease and a prominent dip occurs at 8~AGeV beam
energy. The development of the dip in  $p_x^{\rm dir}$ is directly 
connected to the onset of particle production, as measured by the inelasticity.
At 15~AGeV beam energy, the inelasticity decreases again and the directed flow
increases until a beam energy of 40~AGeV. From there on, $p_x^{\rm dir}$  
stays constant towards the top SPS energy. 

For comparison, simulations with an energy independent elastic and isotropic nucleon-nucleon 
cross section of 40~mb are shown in Fig. \ref{exc.pxdir} (open squares). 
Here one clearly observes a strong
and monotonous increase of the directed flow with energy.

\begin{figure}
\vspace*{-.7cm}
\par \centerline{\resizebox*{!}{0.37\textheight}
{\includegraphics{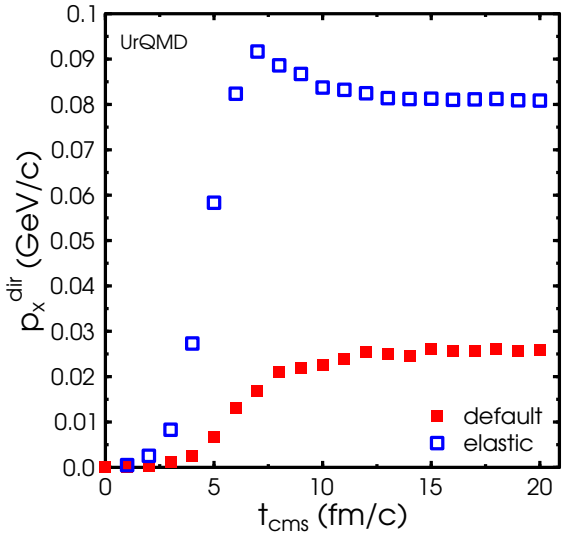}}} \par{}
\caption{Time evolution of the directed flow of nucleons 
at the softest point ($E_{\rm lab}=8$~AGeV). Open squares: only isotropic elastic 
interactions. Full squares: full elastic and inelastic collision term. 
\label{time}}
\end{figure}

While both approaches, hydrodynamics with a phase transition to a QGP or the dynamical 
softening of the EoS presented here, do yield a minimum in the flow excitation function, 
the time evolution of the flow pattern is different.
In scenarios with QGP phase transition, the development 
of a nucleonic anti-flow is visible in the time evolution of the directed flow: with QGP, 
the directed flow at the softest point increases in the early stage 
of the collision, but then suddenly turns over and decreases as the 
anti-flow sets in~\cite{brachmann}.
In Fig. \ref{time}  the time evolution of the directed flow is studied 
in the transport model. Here, the directed flow increases monotonously with time until its
final saturation value.  In both studied scenarios, 
i.e. with the full collision term  included (full squares) and 
in the simulation with only  elastic collisions (open squares), no local 
maximum in the time evolution of $p_x^{\rm dir}$ is present. 
Investigating the directed flow at midrapidity  
shows that in the present model, the minimum in the directed flow
is {\it not} due to a counter-acting anti-flow of nucleons at central 
rapidities as discussed in~\cite{brachmann,Csernai:1999nf}.
Thus, the observation or non-observation of the anti-flow of nucleons 
near central rapidities is crucial
to distinguish hadronic dynamics from  QGP formation.
Unfortunately, up to now all flow measurements at relativistic energies cover only
the range outside $y/y_{cm}\pm 0.3$ \cite{Csernai:1999nf}. The quantitative 
value and the sign of the flow near midrapidity remains undetermined.
Additional information about the origin of the softening of the EoS might also be obtained
by studying the variation of the scaled elliptic flow with centrality at 
fixed energy~\cite{Sorge:1998mk}.
The low energy runs of the NA49 experiment (beam energies of 
20~AGeV, 30~AGeV and 40~AGeV) might be able to explore these 
phenomena in detail \cite{addendum}.
The softening of the equation of state, and the minimum of the directed flow will also
be accessible by the future GSI accelerator \cite{gsifuture}.

In summary, we have demonstrated that a string hadronic model is able to
yield a non-monotonic directed flow excitation function.
In the present model this is due to a dynamical change (softening) of
the equation of state due to a transition to resonance rich matter in line, with 
the onset of particle production.
Irrespective of quantitative uncertainties, we conclude that a  {\it minimum} 
in the  $p_x^{\rm dir}$ excitation function is {\it not} a clean signal for a 
transition from hadron to parton matter in nuclear collisions.

The authors would like to thank  Horst St\"ocker for fruitful and enlightening 
discussions.
M.B. is supported by the region Pays de la Loire. 
M.B. thanks L.P. Csernai and D. R\"ohrich for kind hospitality in Bergen,
were part of this work had been performed. 
This project acknowledges 
support of the Bergen Computational Physics Laboratory in the framework 
of the European Community - Access to Research Infrastructure action 
of the Improving Human Potential Programme.

\end{document}